# Magnetically Tunable Organic Semiconductors with Superparamagnetic Nanoparticles

Rugang Geng[1#], Hoang Mai Luong[1#], Minh Thien Pham[1], Raja Das[2,3], Kristen Stojak Repa[2], Joshua Robles-Garcia[2], Tuan Anh Duong[3], Huy Thanh Pham[3], Thi Huong Au[4], Ngoc Diep Lai[4], George K. Larsen[5], Manh-Huong Phan[2], Tho Duc Nguyen[1]*

[1]Department of Physics and Astronomy, University of Georgia, Athens, Georgia 30605, USA
[2]Department of Physics, University of South Florida, Tampa, Florida 33620, USA
[3]Phenikaa Research and Technology Institute, Phenikaa University, Hanoi, Vietnam
[4]Laboratoire de Photonique Quantique et Moléulaire, UMR 8537, Ecole Normale Supérieure de Cachan, Centrale Supélec, CNRS, Université Paris-Saclay, Cachan, France
[5]National Security Directorate, Savannah River National Laboratory, Aiken, South Carolina 29808, United States
[#]These authors contributed equally to this work.

*Email address: ngtho@uga.edu

**Abstract:** (maximum 250 words)

Magnetic nanoparticles (MNPs) exhibiting superparamagnetic properties might generate large magnetic dipole-dipole interaction with electron spins in organic semiconductors (OSECs). This concept could be considered analogous to the effect of hyperfine interaction (HFI). In order to investigate this model, $Fe_3O_4$ MNPs are used as a dopant for generating random hyperfine-like magnetic fields in a HFI-dominant π-conjugated polymer host, poly(2-methoxy-5-(2-ethylhexyloxy)-1,4-phenylenevinylene) (MeH-PPV). The magnetoconductance (MC) response in organic light emitting diodes made by MeH-PPV/MNP blends is used to estimate the effective hyperfine field in the blends. Firstly, we find that the shape of the MC response essentially remains the same regardless of the MNP concentration, which is attributed to the similar functionality between the nuclear spins and magnetic moments of the MNPs. Secondly, the width of the MC response increases with increasing MNP concentration. Magneto-optical Kerr effect (MOKE) experiments and micromagnetic simulation indicate that the additional increase of the MC width is associated with the magnetization of the blend. Finally, the MC broadening has the same temperature dependent trend as the magnetization of the MNPs where the unique effect of the MNPs in their superparamagnetic and ferromagnetic regimes on the MC response is observed. Magneto-photoinduced absorption (MPA) spectroscopy confirms that the MC broadening is not due to defects introduced by the MNPs but a result of their unique superparamagnetic behavior. Our study yields a new pathway for tuning OSECs' magnetic functionality, which is essential for organic optoelectronic devices and magnetic sensor applications.

# 1. Introduction

Quantum technologies are aimed at exploiting genuine quantum features of systems for use in practical devices. For practical applications of such technologies, electron spins with long lifetimes are necessary for coherent manipulation of spin ensembles and molecular quantum spintronics. Recently, there has been growing interest in using OSECs as quantum materials.[1-3] OSECs are composed of light-elements that possess weak spin-orbit coupling (SOC) and HFI in spins of π-conduction electrons. Therefore, the spin lifetimes in OSECs have been shown to be incredibly long. These long spin lifetimes enable applications in spin-based devices, including qubits in quantum computers, organic spin valves, and spin transistors.[2, 4-10] In such devices, strong spin interaction in the material quenches the long spin coherent time and therefore needs to be suppressed.

On the other hand, many OSECs exhibit intrinsic and large effects of magnetic field in their conductive and luminescent properties, which is also of interest.[11-12] This is commonly known as the magnetic field effect (MFE), and strong spin interactions appear to enhance MFE. It is generally believed that HFI and SOC are the two main spin interactions in OSECs, causing the coherent spin mixing between the singlet and triplet polaron pairs with the same or opposite charge.[13-17] The MFE has been thought to originate from the effects of magnetic fields on the singlet-triplet interconversion rate of loosely bound polaron pairs whose separation is about 1 nm. In such large polaron separations, the contribution of exchange interactions, SOC, and dipolar interactions on the spin mixing can be neglected.[18] Some observable examples of the MFE include organic magnetoresistance (OMAR) in OLEDs, MPA and magneto-photoluminescence in OSEC thin films, and spin dynamics in various chemical/biological reactions.[13, 15, 17, 19-24] The MFE is usually found to be dominant at room temperature in relatively small applied magnetic fields compatible to hyperfine field, making it promising for magnetic field sensing applications down to nT resolution.[25-28]

The presence of HFI in conventional semiconducting polymers has been experimentally proven to be a dominant spin interaction causing MFE.[13-15, 29] This has recently been confirmed by a multi-frequency electron paramagnetic resonance study on OLEDs, where it is found that the SOC affects spin mixing and MFE only when the applied fields exceed 100 mT.[30] Several theoretical models based on the effect of HFI on either spin/electron hopping transport or the intersystem crossing rate of loosely bound polaron pairs have been proposed to effectively explain the MFE in the weak applied magnetic field regime (typically less than 100 mT).[22, 31-34]

Recently, there has been interest in coherent spin manipulation via controlling of the HFI by several state-of-the-art methods, including isotope exchange with different nuclear spins and pulsed electron-nuclear double resonance (ENDOR) spectroscopy.[13-15, 30, 35-38] The former method chemically replaces particular atoms in the molecule by atoms with different nuclear spins.[13-14, 35-37] In conventional conducting polymers, the presence of new isotopes, e.g. $^2$H and $^{13}$C, significantly modified the HFI strengths of hydrogenated OSECs.[13-14, 36] Although this method uniformly manipulates the HFI strength at specified locations in the molecule, it does not offer dynamical control of the interaction. On the other hand, the ENDOR spectroscopy is primarily directed to study the magnetic interactions of the unpaired electron spin with the spins of magnetic nuclei (or HFI) by employing the combined electron paramagnetic resonance (EPR) and the nuclear magnetic resonance (NMR) techniques. While the ENDOR technique is powerful, such an atomic level spin-manipulation technique is often expensive, complex, and therefore not a tabletop apparatus to be widely applied in controlling spin dynamics in OSEC-based devices, especially spin dynamics in organic spin valves.

Most recently, Wang et al.[39] and Cox et al.[40] have demonstrated a method of introducing a randomly localized magnetic field by using the magnetic fringe field of an additional ferromagnetic thin film positioned in the proximity of the OSEC film. Although the magnetic fringe field unambiguously interferes with the hyperfine field to distort the OMAR response, the line shape of the original OMAR response caused by HFI changes significantly. Since the magnetic fringe field in such devices depends on the domain wall switching of the ferromagnetic film and is asymmetrically introduced from one side of the OSEC film, this method generates a biased and strong fringe field, and hence, provides a much smaller degree of field randomization when compared with the hyperfine field.

A decade ago, Cohen theoretically proposed that superparamagnetic MNPs could yield a magnetic dipole-dipole interaction with electron spins, which is analogous to the HFI effect.[18] The magnetic dipole direction of a superparamagnetic MNP experiences quantum fluctuation in ambient temperature, and therefore, induces a quasi-static magnetic field that is randomly oriented on the time scale of microseconds, depending on its size, shape and composition.[41] MNPs of several nanometers in size can also generate uniformly random magnetic fields in the same range of the hyperfine field (several mT).[20, 42] In contrast to the magnetic moment of nuclear spins, the magnetic moment of MNPs in the superparamagnetic regime is easily manipulated by a relatively weak external magnetic field.[41] Therefore, it is possible to manipulate the spin dynamics in OSEC devices by blending these MNPs into the organic emissive layers. We note that incorporation of iron oxide MNPs into the absorption layer of organic solar cells has resulted in enhanced power conversion quantum efficiencies,[43-44] and the use of iron oxide MNPs in hole-injection layers[45-46] or ferromagnetic nanoparticles at an electrode of OLEDs[47-48] has boosted electroluminescence quantum efficiencies (ELQE). However, a clear understanding of the effect of MNPs on spin dynamics of the emissive layers in OSEC devices is lacking. In particular, how magnetic fields induced from MNPs in the two different magnetic regimes (superparamagnetic vs. ferromagnetic) impact OMAR and MPA responses in OSEC films/devices has remained an open question.

In this paper, we investigate the influence of the magnetic fields induced from $Fe_3O_4$ MNPs on the coherent spin mixing in MeH-PPV based OLEDs. In such a polymer, HFI has been proven to play a dominant role in the magneto-transport.[13, 15, 17] By blending the MNPs of different concentrations in the MeH-PPV solution, OLEDs with various levels of magnetic dopants are fabricated. The OMAR effect in OLEDs is measured at different temperatures, crossing the superparamagnetic to ferromagnetic (blocked) transition of the MNPs to gain new insights into the coherent spin mixing in the blends. The effective hyperfine magnetic field in each device is extracted from the half width at half maximum (HWHM) of the OMAR response from which the random field induced by the MNPs is estimated. The induced field versus MNP concentration is also studied by the MOKE spectroscopy and micromagnetic simulation. As control experiments, MPA spectroscopies of the MeH-PPV/MNP and MeH-PPV/Au blend films are investigated for comparison. In all cases, OMAR and MPA are taken within the range of 100 mT to avoid the effect of SOC, and other spin mixing mechanisms such as $\Delta g$ and triplet-triplet annihilation.[30, 49-50]

## 2. Experiment

Iron oxide ($Fe_3O_4$) nanoparticles and MeH-PPV polymer were purchased from NN-Labs LLC and American Dyes Source Inc, respectively. The $Fe_3O_4$ nanoparticles (with a diameter of about 5 nm) capped with oleic acid ligands (~30% by weight) were dispersed in organic solvent toluene as received. The acid groups are chemically bounded to the surface of the particles while the carbon backbones allow the particles to be separated and distributed uniformly in toluene. MeH-PPV polymer was dissolved in toluene with a concentration of 5mg/ml for 24 hours using magnetic stirring bars. When the polymer was well dissolved, the stirring bar was removed from the solution and the MNPs in toluene were added to the polymer solution to get a desired concentration by weight. Finally, the mixture was shaken for 24 hours using a digital vortex mixer to ensure a homogeneous distribution of MNPs.

The device fabrication started with patterning indium tin oxide (ITO) as bottom electrodes. A hole transport layer, poly(3,4-ethylenedioxythiophene) polystyrene sulfonate (PEDOT:PSS, purchased from Ossila) was spin-coated onto a ITO substrate with 5000 rpm ( ~40 nm thick) for 1 minute at ambient atmosphere, and then the PEDOT:PSS film was baked on a hotplate at 120 °C for 2 hours in the nitrogen-filled glovebox. All other fabrication steps are processed inside a nitrogen glove box where the oxygen and water density are less than 0.1 ppm. The blended solution was spin-coated on top of PEDOT:PSS with a spinning speed of 2000 rpm that yields ~100 nm film thickness. The 50 nm calcium was then deposited using a thermal resistive boat followed by 50 nm aluminum deposition using E-beam evaporation in a high vacuum chamber with a pressure of $10^{-7}$ torr. The final device structure is ITO/PEDOT:PSS/blend(MeH-PPV:MNPs)/Ca/Al (see **Fig. 2a**) with an effective area of 1 $mm^2$. After the fabrication, the device was mounted into a cold finger of an optical cryostat with the temperature range of 10 K to 420 K. The cold finger is placed in between two electromagnet poles with magnetic field $B$ up to 300 mT. The device was driven at a constant voltage for a certain designed current density using a Keithley 2400 source and the electroluminescence (EL) intensity from OLEDs was recorded by a Si photodetector. The magneto-conductivity (MC) and magneto-electroluminescence (MEL) were simultaneously measured while sweeping the applied magnetic field $B$ (see **Fig. 2a**). MC and MEL are defined as:

MC=[I(B)-I(B=0)]/I(B=0), MEL=[EL(B)-EL(B=0)]/EL(B=0), where I(B) and EL (B) are device current and EL intensity at applied magnetic field B.

A longitudinal MOKE setup[51] was used to detect the magnetization of MNPs blended in the transparent SU8 2005 photoresist rather than the MeH-PPV polymer, in order to reduce the absorption of the MeH-PPV at the probe wavelength. The MNPs were mixed in SU8 2005 photoresist solution (with a viscosity of 45 cSt) with a certain concentration for 10 days. The SU8 thin films with different MNP concentrations were obtained by spin-coating the nanocomposite solutions on cleaned SiO2/Si substrates with a spin-speed of 2000 rpm resulting in the film thickness of ~ 12 μm.[52] To get a sufficiently high measurement resolution, an ultra-low noise laser diode with 633 nm wavelength and intensity noise of ~ 0.01% and high extinction linear polarizers (>$10^5$ at 633 nm) were used as a probe light and polarizer/analyzer, respectively. In addition, the rotation signal was recorded by a low noise Si detector using the photoelastic modulation (PEM from Hinds Instruments) technique with a 2.405 radian retardation setting for the phase sensitive detection methodology (with a modulation frequency, $2f = 100$ kHz).

## 3. Results and Discussion

**Figure 1a** shows a transmission electron microscope (TEM) image of magnetite particles with an average diameter of 5.5 nm, while the TEM image in **Fig. 1b** shows a quite uniform dispersion of 0.8% MNPs on the MeH-PPV polymer host. The materials were drop-cast on Cu meshes. **Figure 1c** shows the magnetic hysteresis loops of the MNPs at 300 K and 5 K measured by a vibrating sample magnetometer (VSM). The magnetic hysteresis (*M-H*) loops indicate that the nanoparticles are in the superparamagnetic state at 300 K (remanence magnetization $M_r$ ~ 0; coercive field $H_C$ ~ 0), while they are in the ferromagnetic (blocked) state at 5 K with $M_r$ ~ 4.0 emu/g and $H_C$ ~ 6 mT. The insets show enlarged portions of the low-field *M-H* loops, respectively. As compared to bulk Fe3O4 ($M_{S,bulk}$ ~ 90 emu/g), the reduction of the saturation magnetization of the MNPs ($M_S$ ~ 30 emu/g) could be related to the presence of a spin-glass-like shell, also known as the magnetic dead layer, surrounding the nanoparticle, which is composed of disordered spins and does not contribute to the magnetization. For spherical nanoparticles with diameter $d$~5.5 nm, the saturation magnetization is calculated as:[53]

$$M_s = M_{s,bulk}\left(1 - \frac{2D}{d}\right)^3 \qquad (1)$$

with $M_{s,bulk}$ ~ 90 emu/g.[54] Using **Eq. 1** and the $M_s$ value (~30 emu/g at 5 K) in **Fig. 1b**, we can derive out the thickness of the spin-glass-like shell Δ ~ 0.84 nm. The obtained value is similar to that reported previously for Fe3O4 MNPs of similar size.[53]

**Figure 2a** shows the schematic structure of the MeH-PPV/MNP blend based OLEDs. The current-voltage (IV) and EL-V characteristics of the devices at different MNP concentrations (wt. % up to 1.5%) are presented in **Figs. S1a-e**. In general, the onset voltage for current density and EL of all devices are at about 2.4 V and the presence of MNPs quenches the EL intensity (the inset of Fig. S1e, and **Fig. S1f**). For the MNP concentration of 1.5 %, an extremely weak EL intensity was observed (**Fig. S1e**). The EL quenching in the device is probably due to the presence of strong absorption of MNPs at visible light, and the presence of the organic insulating ligands. **Figure 2b** shows the normalized MC response of the OLEDs with different concentrations of $Fe_3O_4$ MNPs in the MeH-PPV/$Fe_3O_4$ blend at 300 K with an applied field range of 100 mT with current density of ~ 80 µA/mm$^2$ at B=0 mT. At this field range, magnetic moments of the MNPs are not fully aligned (see **Fig. 1b**). Consequently, the normalized MC and MEL responses of the pristine OLED at the field up to 200 mT shown in **Fig. S2c&d** saturate much faster than that in OLEDs with MNPs. The inset of **Fig. 2b** is a zoomed-in view of the HWHM for MNP concentration up to 1%, which clearly shows that the higher MNP concentration causes a larger HWHM of MC. Since the MC at 1.5% MNP concentration is quite noisy for directly extracting the HWHM (see **Fig. S2a**), we smoothed this MC response using a double Lorentzian function and obtained the MC width of $(16.5 \pm 0.5)$ mT from the fitted curve.[55] A similar MC width variation is also observed in the MEL response, as shown in **Fig. 2c**. We note that the MEL response at 1.5% MNP concentration is not reliable due the ultralow EL intensity at the 80 µA current (see **Fig. S2d**). Remarkably, the line shape of the MC and MEL responses essentially remains the same regardless of the dopant concentration in **Figs. 2b&c**. This property is attributed to the common functionality between the nuclear spins and the magnetic moments of the MNPs. The inset of **Figure 2d** shows that with increasing the concentration of MNPs, the HWHM of the MC and MEL increases linearly up to 0.8%. However, the behavior increases much faster as the concentration rises above 0.8%. The corresponding magnitudes of the MC and MEL decrease with increasing MNP concentration, as seen in the original MC and MEL results (see **Fig. S2a&b**). The reduction of the MC and MEL magnitudes is associated with the reduction of EL intensity, the important ingredient for the OMAR effect.[11]

We note that the width of the OMAR response is an effective measure of the inhomogeneous local magnetic field strength, $\boldsymbol{B}$ inside the OSECs.[17] Within a pristine MeH-PPV film, the HWHM is mostly due to the local hyperfine field generated by the randomly oriented hydrogen nuclear spins.[13, 15, 54, 56] In a MeH-PPV/MNP blend, the local magnetic field arises from two randomly oriented sources: hyperfine field $\boldsymbol{B_N}$ caused by nearby hydrogen nuclear spins, and the induced magnetic field $\boldsymbol{B_I}$ of the MNPs. By assuming that the effective local magnetic field is a superposition as $\boldsymbol{B} = \boldsymbol{B_N} + \boldsymbol{B_I}$, the width of the MC response corresponds to the case $\boldsymbol{B_N} \uparrow\uparrow \boldsymbol{B_I}$. We can simply estimate the effective induced magnetic field by subtracting the HWHM of 0% MC from that of x% MC, which is defined as ΔHWHM = HWHM(x%) – HWHM (0.0%). **Figure 2d** shows the effective induced magnetic field, ΔHWHM from the MNPs at different concentrations that is extracted from the HWHM of MEL and MC response. ΔHWHM linearly increases with MNP concentration below 0.8% and increases exponentially for larger MNP concentrations. In order to confirm that the ΔHWHM broadening is associated with the induced magnetic field from the MNPs, we performed the MOKE spectroscopy of MNPs blended in the transparent SU8 2005 photoresist. The MOKE responses in 12 µm SU8/MNP(%) films are shown in **Fig. S3**. After 100 times averaging the MOKE signal has a resolution of approximately 3 µdeg (**Fig. S3**, **inset**). In general, the Kerr rotation angle is not a linear function of the MNP concentration. The magenta open circles in **Fig. 2d** with specific error bars show the saturated Kerr rotation angle/magnetization of the films extracted from the MOKE spectra. Notably, the magnetization of the films matches well with the randomly induced field extracted from the OMAR measurements. This validates the hypothesis that the additional field characteristic in the film measured by OMAR indeed arises from the MNPs.

In order to further investigate the effect of MNPs on OMAR response, the magnetic moments of the MNPs are modified by controlling their temperature. **Figure 3a** shows that the effective induced magnetic field increases with decreasing temperature, which generally agrees with the temperature-dependent magnetization of MNPs as shown in **Fig. 3b** where the magnetization was measured in a zero-field-cooling (ZFC) regime.[57] The temperature-dependent MC and MEL responses of pristine MeH-PPV and MeH-PPV/MNP (0.8%) films in the 100 mT field range are shown in **Figs. S4 & S5**, where the change in the MC width in **Fig. 3a** caused by the MNPs is calculated. The temperature dependence of the current density of the pristine and blend devices is shown in **Figs. S6**. In general, the magnetization increases with decreasing temperature, so does MC width. However, at the blocking temperature of ~16 K, the magnetization of ZFC reaches a maximum and then decreases at lower temperatures. Interestingly, the effective induced magnetic field is similarly maximized at a specific temperature, followed by a slight decrease at lower temperatures. The curve resembles the ZFC magnetization behavior of MNPs[57] but the effective blocking temperature $T_B$ has been extracted to be ~ 60 K ($\pm$ 20 K), which is much higher than that extracted from the M(T) curve measured by Physical Property Measurement System (PPMS) (**Fig. 3b**). This interesting phenomenon might be due to the different measurement time scales in the two methods. MC is originated from a non-equilibrium process that has typical time scales in the microsecond range, while it is on the order of 1.0 second for the PPMS.[58-60] In addition, the magnetization measurement by PPMS in **Fig. 3b** is for a cluster of MNPs. If the magnetic interactions between the MNPs are not negligible, they can have a significant influence on the superparamagnetic relaxation. It is worth noting that we did not utilize the temperature-dependent MEL response for extracting the induced magnetic field because the triplet-triplet annihilation might happen strongly at low temperature leading to a negative component of MEL at high magnetic field, which would make the induced field extraction less accurate.[49]

It is described in **Fig. 1b** and **Fig. 3b** that the MNPs experience a magnetic phase transition from the superparamagnetic to ferromagnetic regime at a low temperature of about 16 K. The MNPs might generate interesting induced field dynamics that occur at the scale smaller than several mT, especially at this transition region, and such dynamics within small magnetic field scale might not be captured by the MC studies in the intermediate applied field range. Therefore, the ultra-small MC at magnetic field range of less than 1 mT has been studied to capture this interesting magnetic phase transition.[15] **Fig. 4a** shows the ultra-small MC response of the pristine (0.0% MNPs) OLED and 0.8% MNP based OLED at room temperature where constant bias voltages were applied, corresponding to an electric current of about 80 µA in each device. The magnetic characteristic, $B_{min}$ remains at ~ 0.36 mT in both cases (the inset of **Fig. 4a**). It is surprising that MNPs do not have an effect on the shape of the ultra-small MC, while causing a significant reduction in the MC magnitude at $B_{min}$, $MC_{min}$. It should be noted that the ultra-small MC response depends on temperature and bias voltage.[15] However, the ratio, $MC/MC_{max}$ was found in the literature to be insensitive to the temperature, where $MC_{max}$ is the magnitude of a saturated MC.[15] Our result of the temperature-dependent $MC/MC_{max}$ ratio of a pristine MeH-PPV device shown in **Fig. S7** agrees well with this report. Therefore, any change of the $MC/MC_{max}$ response is confidently attributed to the change caused by the induced magnetic field. **Fig. 4b** shows the temperature dependence of the $MC/MC_{max}$ response of the blended OLED with 0.8% MNPs measured at a current of 80 µA, where the device was cooled down to 10K with zero applied field. The magnitude of the $MC_{min}$ is slightly suppressed at low temperatures due to the significant increase in the magnetization of the MNPs (**Fig. 3b**), while $B_{min}$ is essentially the same when the temperature is above the blocking temperature. This is in agreement with the temperature independent $B_{min}$ in the pristine film-based device (see **Fig. S7**). However, when the temperature goes down to 10K, the MNPs experience the phase transition from superparamagnetism to ferromagnetism. The ultra-small MC shape and hence $B_{min}$ substantially changes. This effect is likely related to the blocked, ferromagnetic state of the MNPs where the applied magnetic field (~1 mT) was not large enough to rotate the blocked magnetic moments, resulting in this unusual MC behavior. The relatively large static field from the ferromagnetic particles might override the hyperfine field at this temperature erasing the effect of the ultra-small MC. In summary, the presence of the superparamagnetic particles in MeH-PPV suppresses the $MC_{min}$ differently at different temperatures but leaves the $B_{min}$ essentially unchanged. A theoretical investigation is necessary to understand the ultra-small MC response under the influence of MNPs in both superparamagnetic and ferromagnetic regimes, which is beyond the scope of the current paper.

Since the MNPs and oleic acid ligands might slightly change the charge injection balance and might induce charge defects in the OLEDs that may also influence the MC response, we perform photo-induced spectroscopy (PA) and MPA of the pristine and blend films. The PA spectroscopy is an experimental technique used to study long lifetime species such as triplet and polaron excitations. With the PA technique, the charge impurity can be sensitively detected.[61] In this experiment, a cw diode laser at 450 nm is used as a pump to promote electrons from the ground to excited states forming excited species such as singlet, triplet states and free charges. A tungsten lamp with broad spectrum is then used as a probe light to detect the excited state density of the long lifetime species. **Figure 5a** shows typical PA spectra of a pristine MeH-PPV thin film and the MeH-PPV blend film with 1% MNP concentration. Their spectrums are almost identical with each other. There is no evidence that significant charge impurities are formed in the MNP blend since their polaron absorption signatures at 0.5 eV and 1.6 eV for the charge impurities in the spectrum are not observed.[61] Next, the blend and pristine MeH-PPV films are irradiated by a xenon-lamp for 2 hours to generate photo-induced native defects in the film. These defects produce substantial long-lived photogenerated polarons that have two characteristic PA bands at 0.5 eV and 1.55 eV in the PA spectrum as described in our previous report.[62] Such photogenerated polarons can form loosely bound polaron pairs whose singlet/triplet ratio can be influenced by HFI and applied magnetic field, providing insight into the nature of local magnetic field. In a magnetic field, $B$, PA($B$) is determined by field dependent photoexcitation density, since the photoexcitation optical cross-section, σ(E) changes negligibly with $B$. Magneto-PA is defined by MPA($B$)≡[PA($B$)-PA(0)]/PA(0).[62] **Figure 5b** shows the normalized and smoothed MPA responses at 1.55 eV of a pristine MeH-PPV film and a film blended with 1% MNPs. The original MPA responses could be found in the supplemental material (see **Fig. S8**). The HWHM of MPA with MNPs (red line) obtaining directly from the figure is 7.9 mT, which is ~1.2 mT broader than the MPA of the pristine film (blue line). These results are consistent with the OMAR response in OLEDs described above. This confirms that the induced magnetic field from MNPs causes the MFE broadening in a similar manner as when $^{13}$C isotope is introduced into the polymer.[36, 63] Finally, since the MPA and MC broadening might arise from an increase of SOC generated at the surface of the $Fe_3O_4$ nanoparticles or/and the ligand effects, we perform a control MPA experiment where $Fe_3O_4$ particles in MeH-PPV are replaced by Au nanoparticles of similar size having the same ligands. The inset of **Fig. 5b** shows the MPA measurement of a pristine film and a film with 1% Au NPs (10 nm diameter) synthesized by our lab. The normalized MPAs of the two films have identical HWHM, indicating that the magnetic response broadening of MC and MPA must come from the induced magnetic field from the MNPs, and the strong SOC from Au has negligible influence on the spin mixing in the material.

To further understand how the induced magnetic field from MNPs behaves as an artificial HFI that affects the spin dynamics in MeH-PPV, we now interpret our results by the general effects of random hyperfine field and induced magnetic field on the spin mixing between singlet and triplet polaron pairs.[15] In a pure MeH-PPV film, spins of polarons precess around their randomly oriented local magnetic field, $B_{N,i}$ (i=1 or 2) from the nuclear spins causing the spin mixing between singlet and triplet polaron pairs (**Fig. 6a**). When MNPs are blended in the films, the local field now comprises of hyperfine field and randomly oriented induced magnetic field from MNPs as $B_i = B_{N,i} + B_{I,i}$ causing stronger spin precession or singlet/triplet polaron pair spin mixing rates (**Fig. 6b**). In general, the polaron spin will precess at a local magnetic field $B$ with a frequency given by

$$\omega_B = \mu_B B g / \hbar \qquad (2)$$

where g is the g-factor of the polaron, $\mu_B$ is the Bohr magneton, and $\hbar$ is the Planck constant. The difference of the precession frequency between two adjacent polarons' spins can be written as

$$\Delta \omega_B = \frac{\mu_B}{\pi \hbar}(B\Delta g + g\Delta B) \qquad (3)$$

The first term $B\Delta g$ is referred to as the $\Delta g$ mechanism, which has been mostly used to explain the magnetic field effect in OSEC blends with large magnetic field and/or in OSEC blends.[50, 64-66] In homogeneous materials like MeH-PPV, we ignore the effect of $B\Delta g$ due to the negligible difference of g-factors between polarons and the small applied magnetic field.[30] The second term $g\Delta B$ is referred to as the $\Delta B$ mechanism, where $\Delta B$ is the difference of magnetic field at two adjacent sites, resulting in different spin precession rates causing the spin mixing between singlet states and triplet states.[18, 40] Without MNPs, the spin mixing rate is proportional to $\Delta B = |\Delta \boldsymbol{B_N}|$, where $\Delta \boldsymbol{B_N}$ is the nuclear field gradient (**Fig. 6a**). It is obvious that if the hyperfine field, $\boldsymbol{B_{N,1}}$ and $\boldsymbol{B_{N,2}}$ are the same, the spin mixing does not occur. With the presence of MNPs, the different spin precession rate between polarons, $\Delta \omega_B$ is proportional to $\Delta B = |\Delta \boldsymbol{B_N} + \Delta \boldsymbol{B_I}|$, where $\Delta \boldsymbol{B_I}$ is induce field gradient (**Fig. 6b**). If $\Delta \boldsymbol{B_I}$ is comparable to $\Delta \boldsymbol{B_N}$, this would significantly enhance the mixing rate between singlet and triplet polaron pairs. This is the principle for spin mixing mechanisms in the loosely bound electron-hole pair model, and the bipolaron model (spin mixing between two polarons of the same kind) for MFE.[13-17]

For estimating the average strength of the induced magnetic field of MNPs in their proximity, we performed a micromagnetic simulation of the induced magnetic field using an open source calculation code, OOMMF (The Object Oriented MicroMagnetic Framework). For simplicity, we assume that MNPs with parallel magnetization are dispersed uniformly in the MeH-PPV host. In our simulation, the material parameters for $Fe_3O_4$ nanoparticles were selected as following:[54, 67] saturation magnetization, $M_s=3.65*10^5$ A/m; anisotropy constant, $K=1.1*10^4$ J/m$^3$; exchange constant $A= 1.2*10^{-11}$ J/m; and diameter of MNP d = 3.8 nm, where the magnetic dead layer of 0.84 nm has been considered. The mesh size was selected as small as 0.5 nm. In addition, we ignore the magnetic interaction between MNPs since the distance between two neighboring MNPs is much larger than the size of MNPs (inset of **Fig. 6d**). As shown in **Fig. 6c**, the induced magnetic field is very strong near the MNP surface and then exponentially decays with the distance. The color bar shows the magnitude of the induced magnetic field in the log scale. The overall average induced field gradients with a 1nm separation can be extracted from **Fig. 6c**. **Figure 6d** shows that the average divergence linearly increases with increasing the particle concentration in the films. It well matches with the linear dependence of ΔHWHM at the concentration below 0.8% (replotted from **Fig. 2d**, inset). Nevertheless, the ΔHWHM increase rapidly when the concentration is above 0.8%, in contradiction to the magnetic simulation. We speculate that since MNPs in the simulation are considered non-interacting, while in reality, some clusters of MNPs may form, leading to the significant enhancement of the induced magnetic field in experiment due to the strong particle-particle interaction at above 0.8% concentration. In addition to MNP interactions, we note that the slope of the field gradients in the simulation may also depend on other parameters, such as the polaron separation distance or particle size distribution. This can result in a lightly larger slope of the field gradients in the experiment than that in the simulation.

**4. Conclusion.**

We quantitatively studied the effect of the induced magnetic field from $Fe_3O_4$ MNPs on the spin dynamics in MeH-PPV based OLEDs. We found that the HWHM of the OMAR and MPA response of MeH-PPV/ $Fe_3O_4$ blend films increases with increasing the concentration of $Fe_3O_4$ due to the enhancement of the local magnetic field. The induced field behaves as a second hyperfine field in super-position with the intrinsic hyperfine field from the hydrogen nuclear spins. The properties of the induced magnetic field as studied by VSM, MOKE, and micromagnetic simulation of the MNPs are in excellent agreement with the MFE results. Control experiments confirm that the observed characteristics are not due to defects introduced by the MNPs, but are instead the result of the uniqueness of superparamagnetic behavior. Our study yields a new pathway for tuning OSECs' magnetic functionality, which is essential to organic optoelectronic devices and magnetic sensor applications.

**Acknowledgement**

This work was supported by the UGA start-up fund (T. D. N.), Savannah River National Laboratory's Laboratory Directed and Development program (SRNL is managed and operated by Savannah River Nuclear Solutions, LLC under contract no. DE-AC09-08SR22470) (T. D. N. and G.K.L.), and the grant from VICOSTONE USA (T. D. N.). M.H.P acknowledges the U.S. Department of Energy, Office of Basic Energy Sciences, Division of Materials Sciences and Engineering under Award No. DE-FG02-07ER46438.

**Figure Captions:**

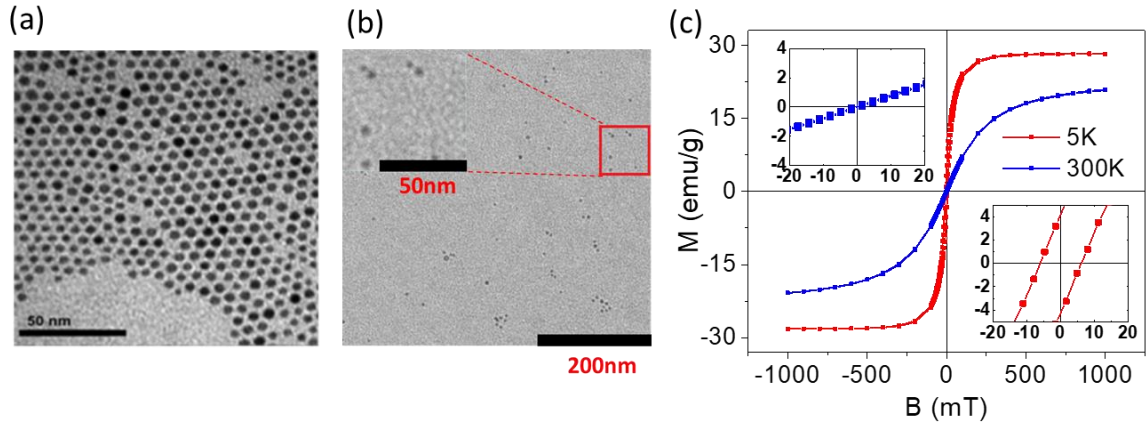

**Figure 1**. (a) Transmission electron microscope (TEM) images of (a) a pure $Fe_3O_4$ film and (b) a composition film made by 0.8% $Fe_3O_4$ concentration on the MeH-PPV host. (b) The magnetic hysteresis M(H) loops of $Fe_3O_4$ at 300 K and 5 K measured by the PPMS. Insets show the low-field portion of the M-H loop taken at 300 K and 5 K.

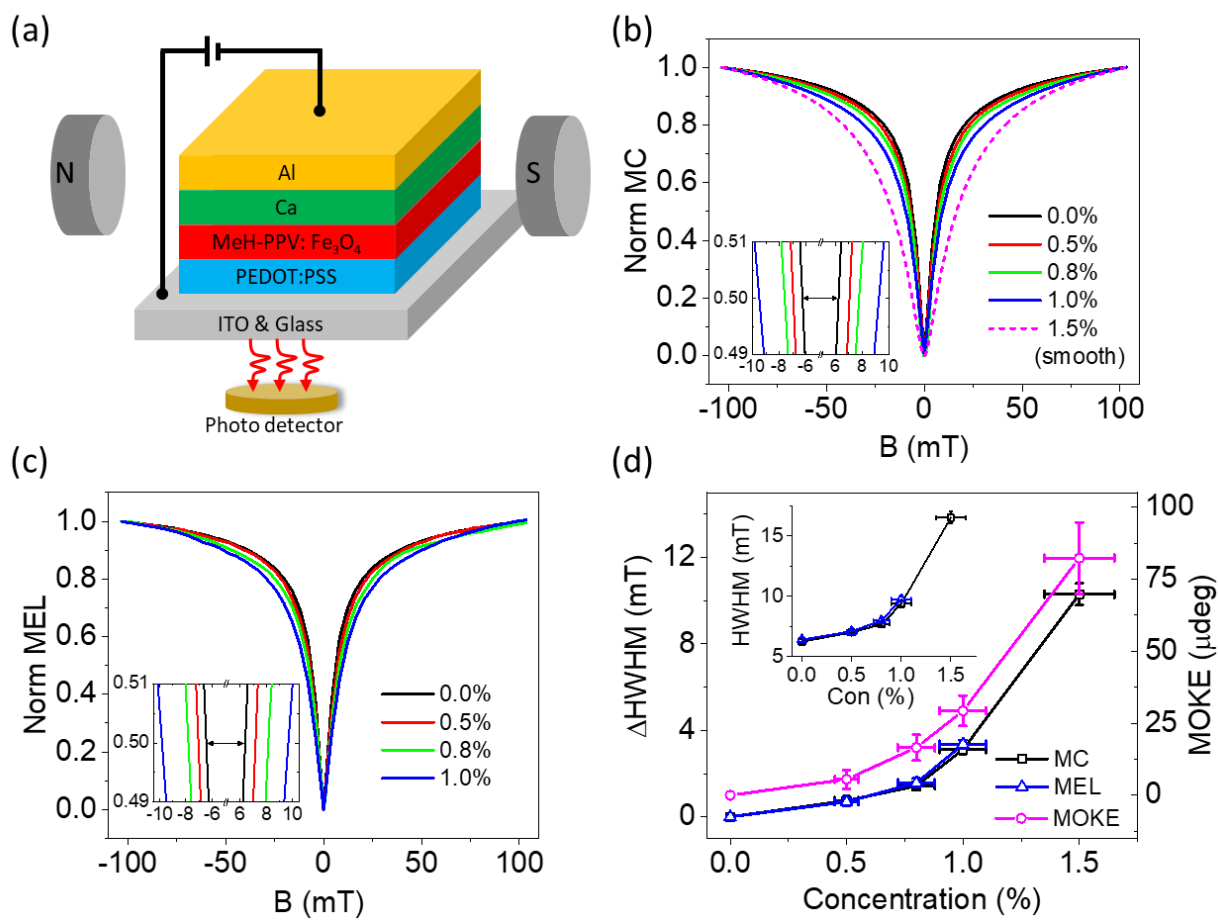

**Figure 2.** (a) Schematic structure of the OLED with an external magnetic field, where the active layer is a blend of MeH-PPV polymer and $Fe_3O_4$ MNPs. Normalized (b) MC and (c) MEL responses in OLEDs at different concentrations of the MNPs (weight ratio) with current ~80 µA at B=0 mT. Insets in (b) and (c) show a zoomed in view of the MC and MEL responses for the MNP concentration up to 1%, respectively. The dash line shows the smoothed MC response of 1.5% MNP concentration (the original data is shown in Figure S2) by a double Lorentzian function. (d) The inset shows the HWHM of the MC and MEL at several concentrations of the MNPs while the main panel shows the induced magnetic field (Δ HWHM) from the MNPs that are extracted from the inset. The error bars for both MNP concentration and HWHM are presented. The magenta open circles with the error bars show the MOKE rotation angles at the corresponding MNP concentrations. The MOKE data is vertically off-set from the Δ HWHM.

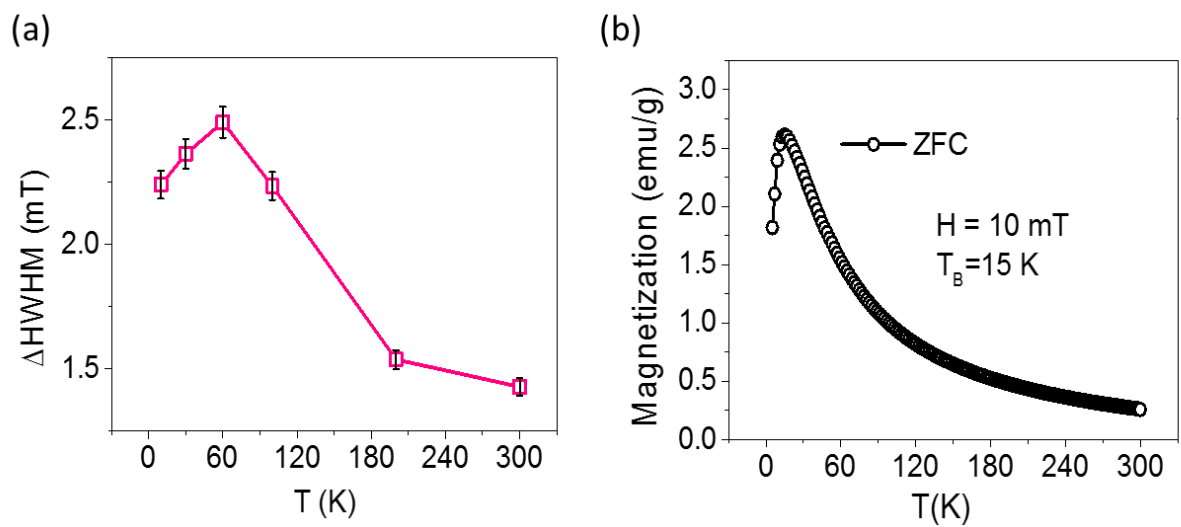

**Figure 3.** (a) The difference in HWHM of MC between pristine (0 %) and 0.8 % concentration-based OLEDs with temperature dependence. The error bars are shown. (b) The temperature-dependent magnetization of the MNPs with zero field cooling (ZFC).

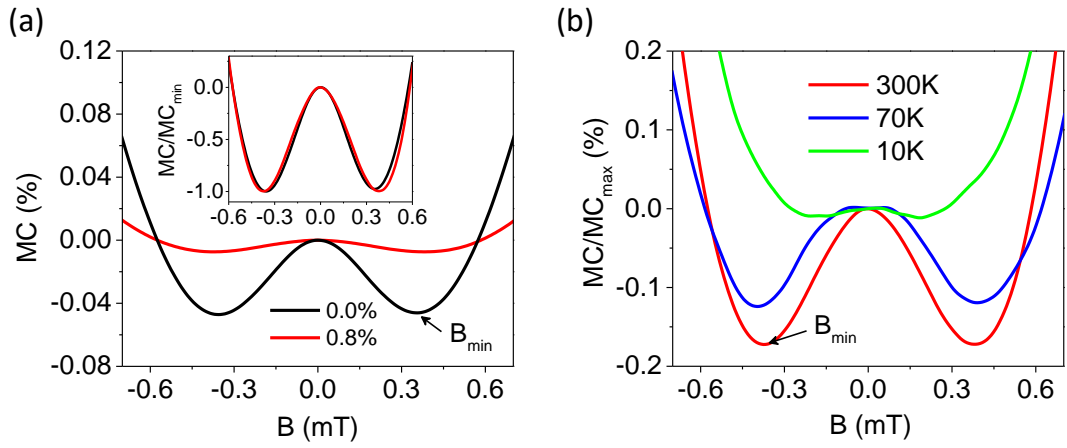

**Figure 4.** (a) Ultra-small field MC response at room temperature with 0 % and 0.8% concentration. Inset shows the normalized MC with the $MC_{min}$. (b) Temperature dependent $MC/MC_{max}$ of 0.8% MNP based OLED.

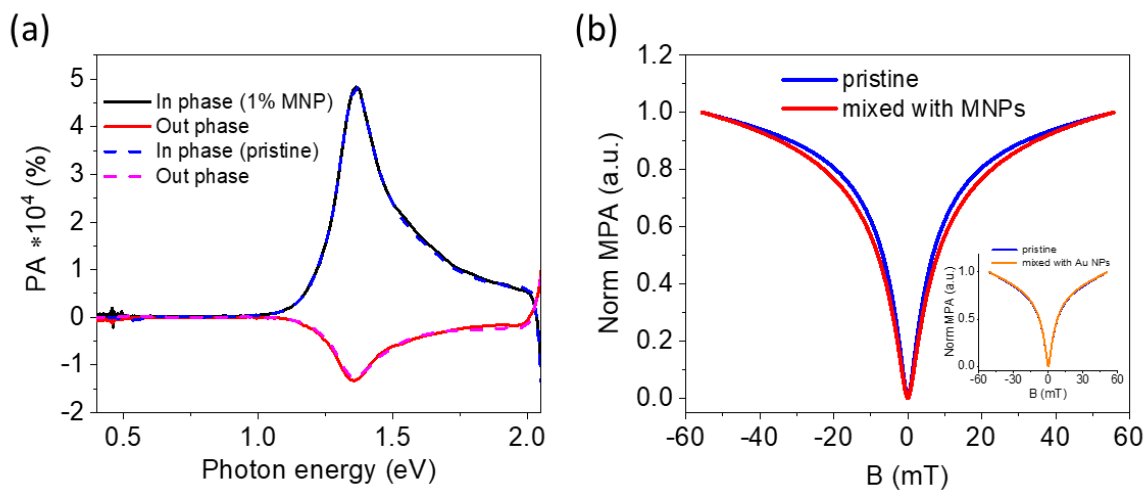

**Figure 5.** (a) PA spectra of a pristine MeH-PPV thin film (dash lines) and a MeH-PPV/MNP(1%) thin film (solid lines). (b) Normalized smoothed MPA response of a pristine MeH-PPV film and the film blended with 1% MNPs taken at 1.55 eV. The inset shows the normalized smoothed MPA response of a pristine film and the film blended with 1% Au nanoparticles.

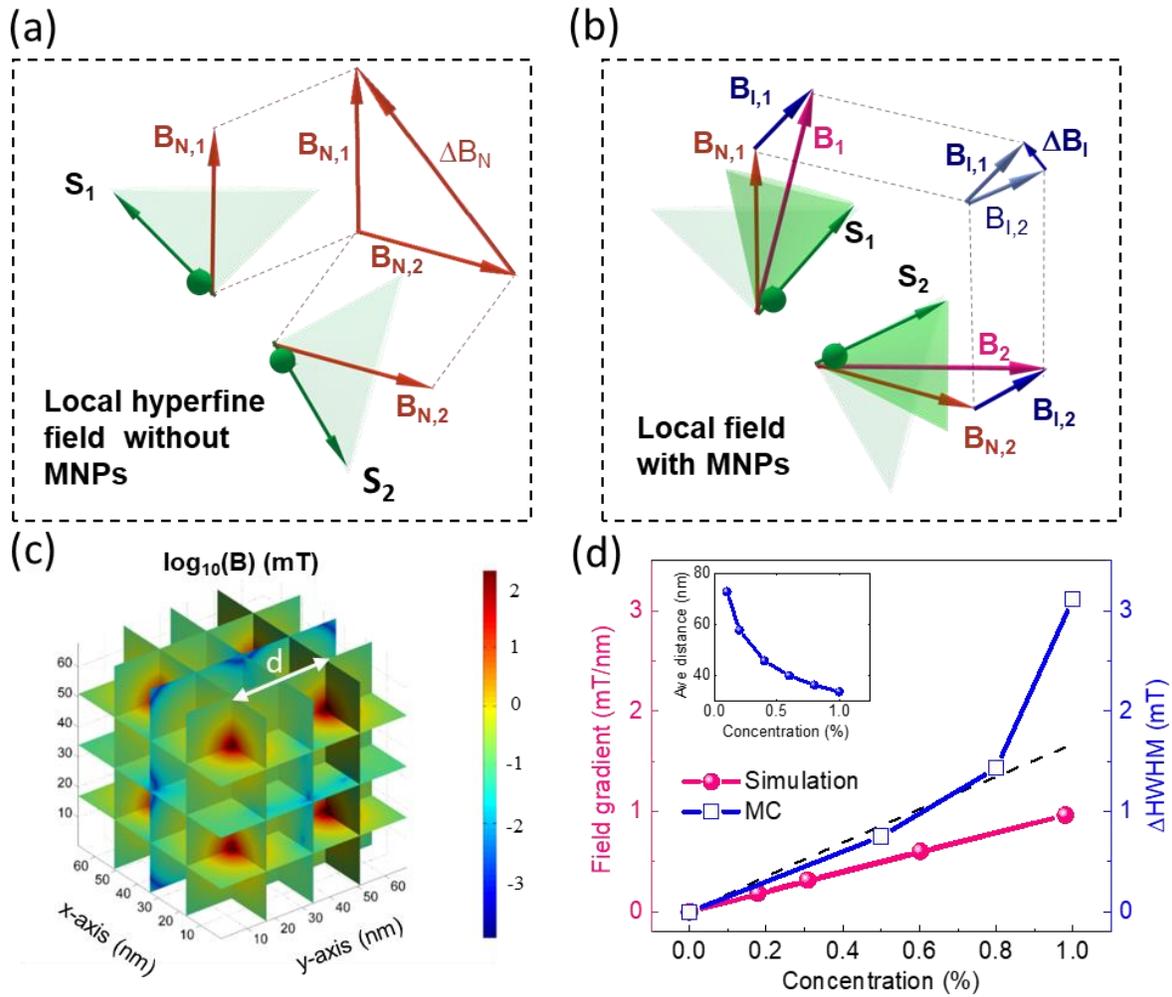

**Figure 6.** Spin precession of two polarons' spins, $S_1$ and $S_2$ in a spin pair causing the spin mixing in the case (a) without and (b) with MNPs. $B_{N,1}$ and $B_{N,2}$ are effective random hyperfine fields while $B_{I,1}$ and $B_{I,2}$ are effective induced magnetic fields at the polaron positions. $\Delta B_N$ and $\Delta B_I$ are the nuclear field and induced field gradients, respectively, seen by the polarons (c) Spatial distribution of the induced magnetic field in a log scale with the unit of mT, where the lattice distance d=33.8 nm (1% MNP) and the diameter of MNP is 3.8 nm. (d) The average of induced field gradients at different NMP concentrations in three dimensions. For comparison, the ΔHWHM of MC versus particle concentration is plotted together with a linear fit at concentrations below 0.8%. The inset shows the average distance d with the concentration dependence.

Supplemental Information

**Magnetically Tunable Organic Semiconductors with Superparamagnetic Nanoparticles**

Rugang Geng[1#], Hoang Mai Luong[1#], Minh Thien Pham[1], Raja Das[2,3], Kristen Stojak Repa[2], Joshua Robles-Garcia[2],Tuan Anh Duong[3], Huy Thanh Pham[3], Thi Huong Au[4], Ngoc Diep Lai[4], George K. Larsen[5], Manh-Huong Phan[2], Tho Duc Nguyen[1]*

[1]Department of Physics and Astronomy, University of Georgia, Athens, Georgia 30605, USA
[2]Department of Physics, University of South Florida, Tampa, Florida 33620, USA
[3]Phenikaa Research and Technology Institute, Phenikaa University, Hanoi, Vietnam
[4]Laboratoire de Photonique Quantique et Moléulaire, UMR 8537, Ecole Normale Supérieure de Cachan, Centrale Supélec, CNRS, Université Paris-Saclay, Cachan, France
[5]National Security Directorate, Savannah River National Laboratory, Aiken, South Carolina 29808, United States

[#]These authors contributed equally to this work.

*Email address: ngtho@uga.edu

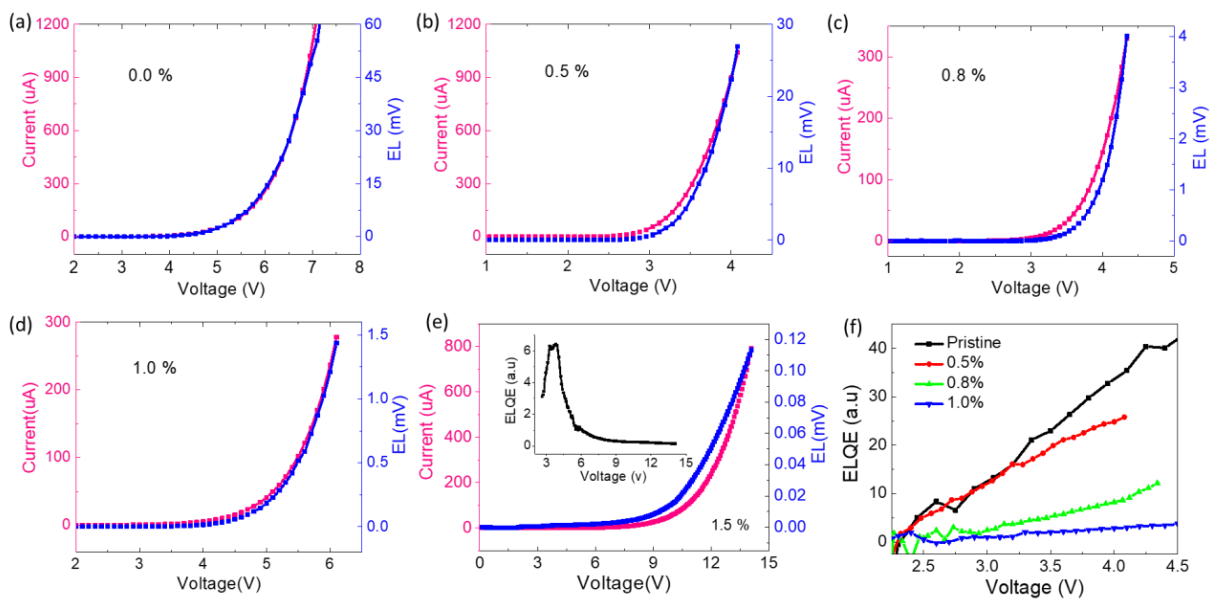

**Figure S1.** The current-voltage (IV) and EL-V characteristics of the devices with (a) 0.0 %, (b) 0.5 %, (c) 0.8 %, (d) 1.0 %, and (e) 1.5% MNP concentration at room temperature, (f) Electroluminescence quantum efficiency (ELQE defined by EL/I) versus applied voltage. The data was processed from parts a-d. The ELQE of 1.5% MNP device is plotted in the inset of (e).

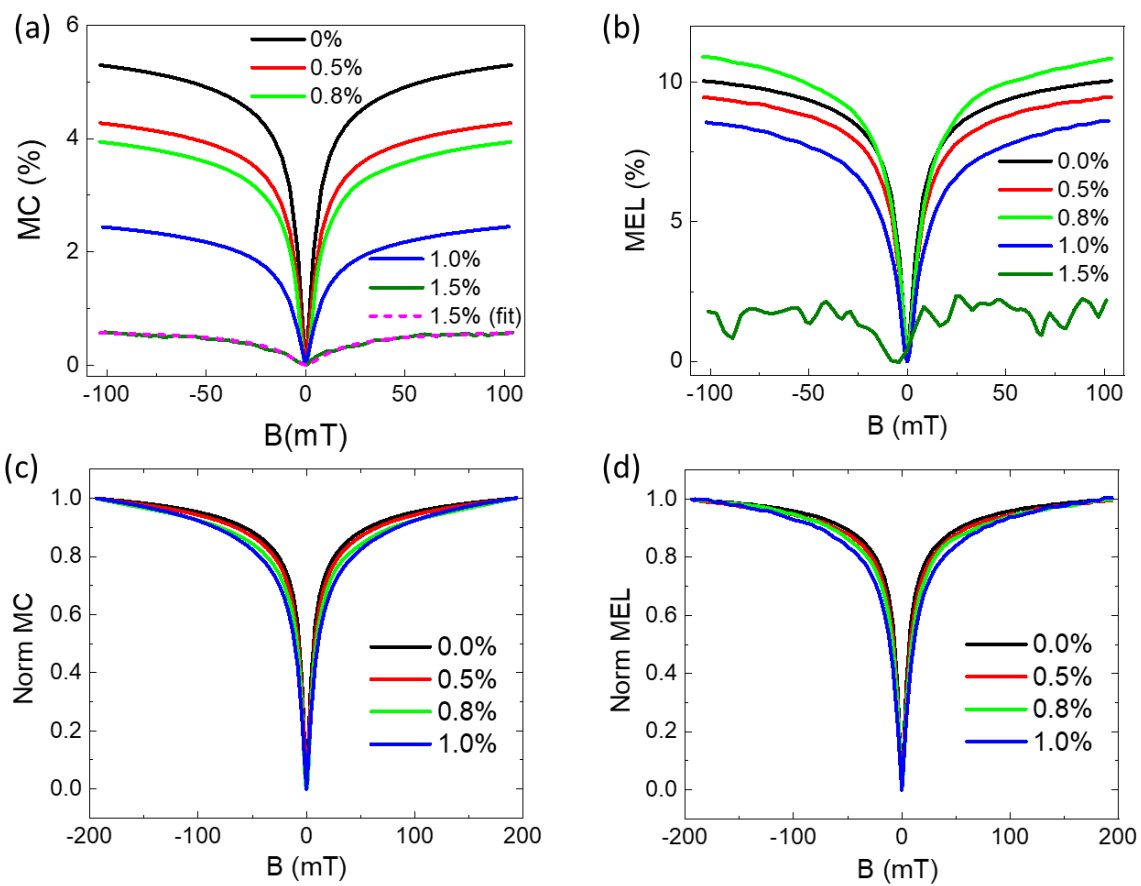

**Figure S2.** Magnetic field effect of the OLEDs with different concentrations of MNPs at 300 K. (a) MC response and (b) MEL response with the concentration dependence with magnetic field range of 100 mT. Normalized (c) MC and (d) MEL with magnetic field range of 200 mT.

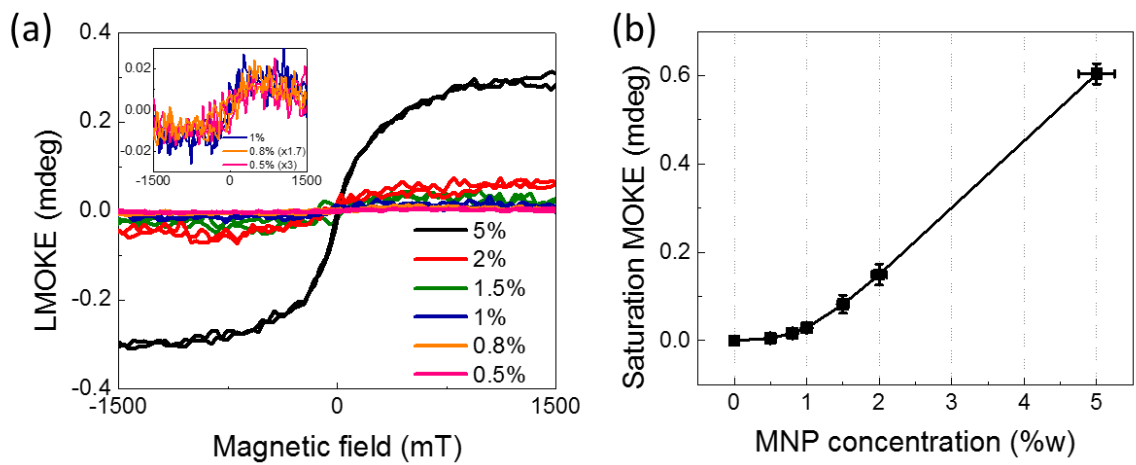

**Figure S3.** (a) MOKE hysteresis curve of SU8/MNP thin film with different concentration of MNPs. Inset: zoom-in of MOKE hysteresis curve with low concentration of MNP of 1% and 0.5%. (b) Saturation MOKE signal versus concentration of MNPs (%w) with error bars.

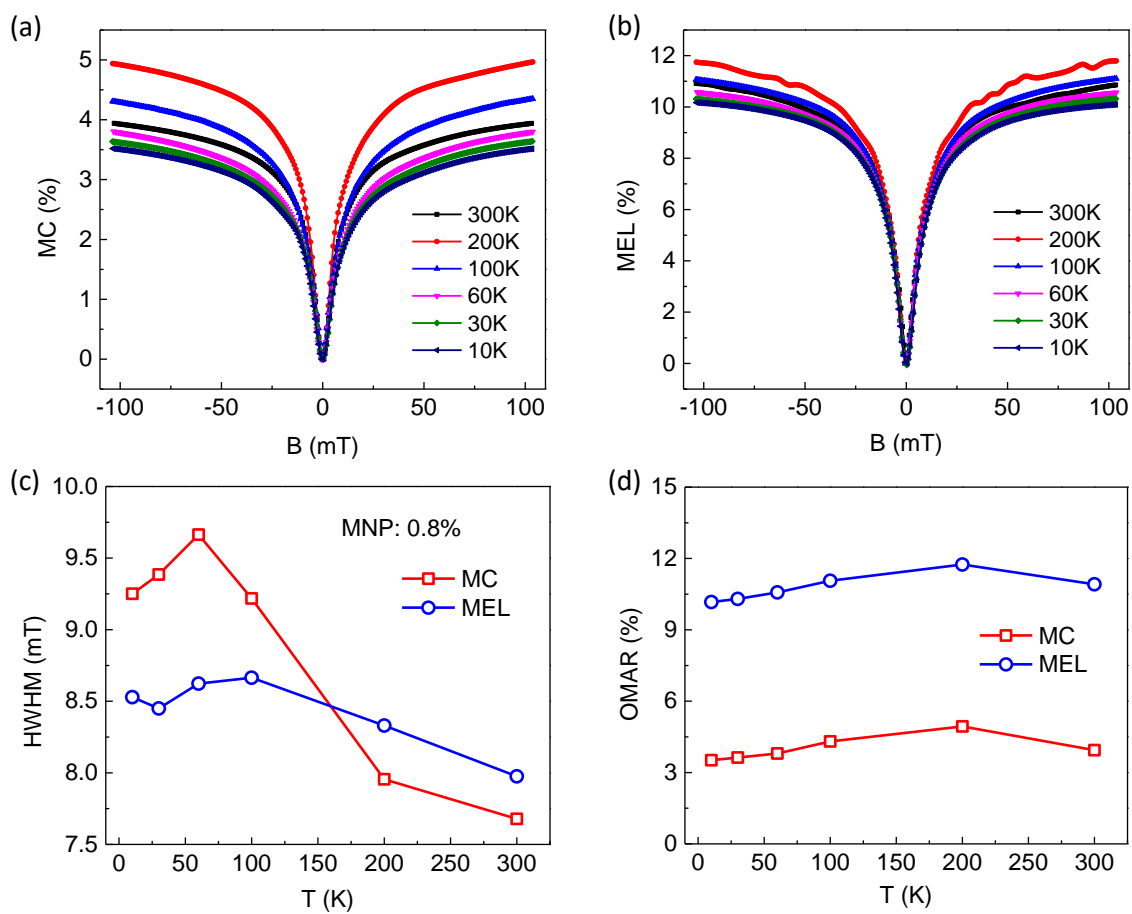

**Figure S4.** Magnetic field effect of the blend OLED with 0.8% MNP at different temperatures. (a) MC responses and (b) MEL responses with temperature dependence. (c) Half-width of half maximum (HWHM) and (d) the magnitude of MC and MEL responses with temperature dependence.

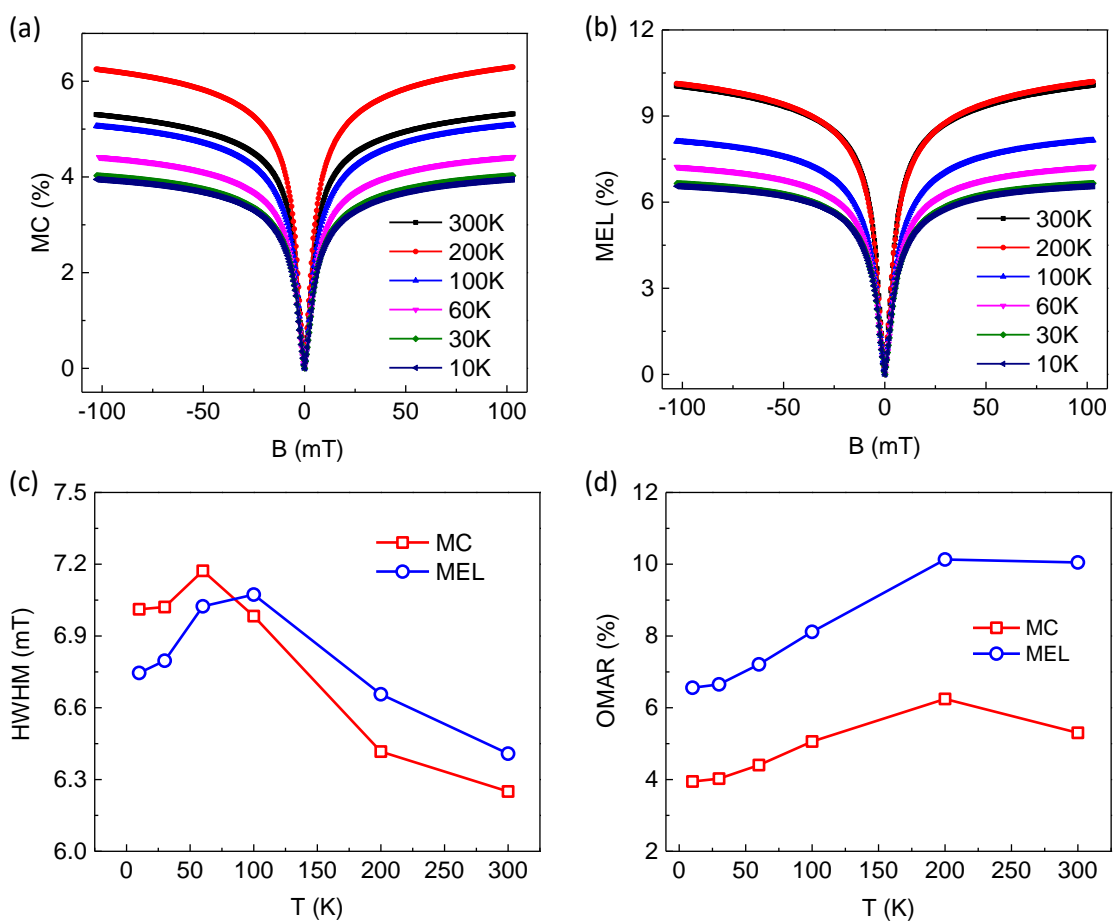

**Figure S5.** Magnetic field effect of the pristine OLED at different temperatures. (a) MC responses and (b) MEL responses with temperature dependence. (c) Half-width of half maximum (HWHM) and (d) the magnitude of MC and MEL responses with temperature dependence.

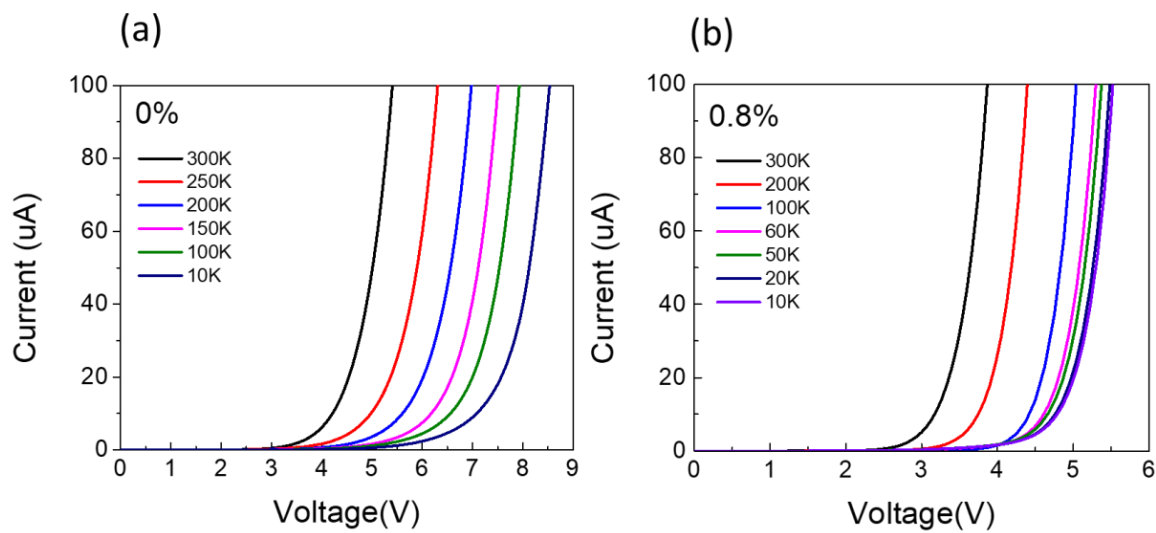

**Figure S6.** Typical temperature dependence of current density in (a) pristine and (b) 0.8% MNP based OLEDs.

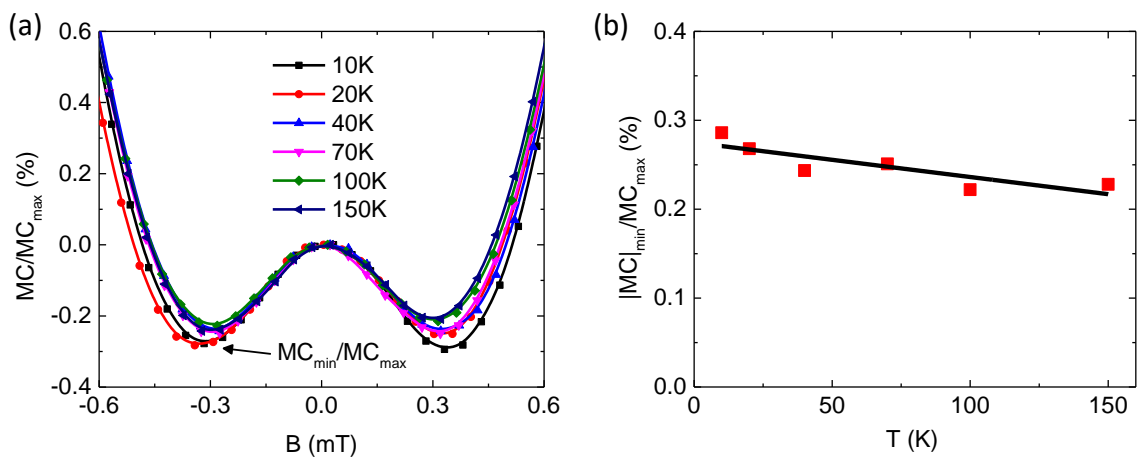

**Figure S7.** MC response of a pristine OLED at the ultra-small magnetic field. (a) Normalized MC response at the ultra-small magnetic field with temperature dependence. (b) $|MC|_{min}/MC_{max}$ at different temperatures, where the data points (red square) is linearly fitted (black line).

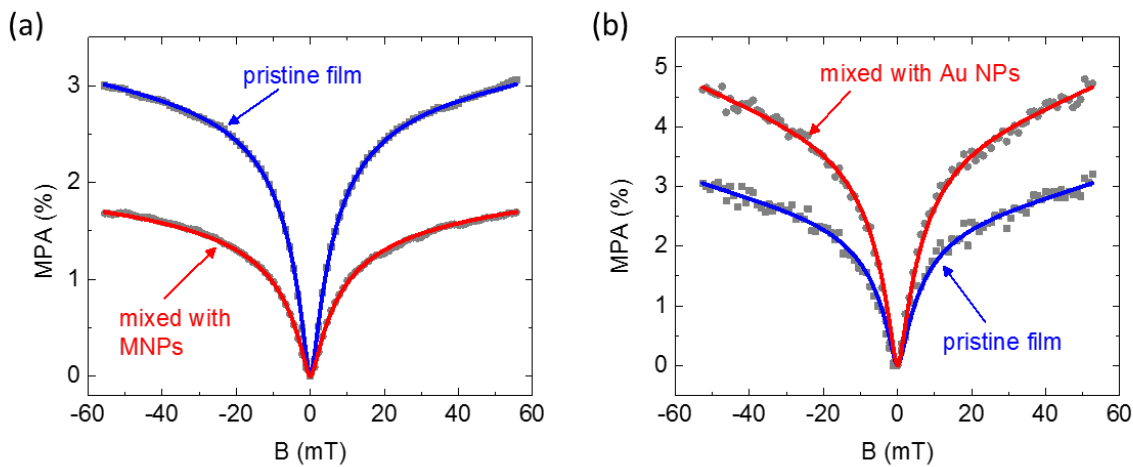

**Figure S8.** Magneto photo-induced absorption (MPA) measurement of the pristine MeH-PPV films and films blended with nanoparticles. (a) MPA of a pristine film and the film blended with 1% $Fe_3O_4$ MNPs. (b) MPA of a pristine film and the film blended with 1% Au nanoparticles (10 nm in diameter). The MPA data are fitted by using a double-Lorentzian function, and the fitting results are the blue and red curves in (a) and (b). The purpose of the fitting is to have smooth curves for the comparison shown in the main text. We did not aim to compare the fitting parameters since it is not the scope of this report. It is worth noting that the MPA magnitude of the film with Au nanoparticles is larger than that in the pristine film, in contrast to the MPA of the film with MNPs. The enhancement of MPA in the Au containing film might be due to the strong localized surface plasmon resonance at the Au nanoparticles.